\documentclass[aps,prl,twocolumn,showpacs,superscriptaddress,floatfix,nofootinbib]{revtex4-1}
\usepackage{graphicx,graphics}
\usepackage{dcolumn}
\usepackage{amsmath,amssymb,amsfonts}
\usepackage{latexsym,verbatim}
\usepackage{bm}
\usepackage[toc,page]{appendix}
\usepackage{color}
\usepackage{ulem}
\usepackage[breaklinks=true,colorlinks,citecolor=blue,linkcolor=blue,urlcolor=blue]{hyperref}
\usepackage{tabularx}
\graphicspath{fig}

\def\be{\begin{eqnarray}}
\def\ee{\end{eqnarray}}

\def\<{\left\langle}
\def\>{\right\rangle}

\newcommand{\beq}{\begin{equation}}
\newcommand{\eeq}{\end{equation}}

\begin{document}
\title{Lattice study of electromagnetic conductivity of quark-gluon plasma in external magnetic field}

\author{N.\,Yu.~Astrakhantsev}
\email[]{nikita.astrakhantsev@phystech.edu}
\affiliation{Physik-Institut, Universit{\" a}t Z{\" u}rich, Winterthurerstrasse 190, CH-8057 Z{\" u}rich, Switzerland}
\affiliation{Moscow Institute of Physics and Technology, Institutsky lane 9, Dolgoprudny, Moscow region, 141700 Russia}
\affiliation{Institute for Theoretical and Experimental Physics NRC ``Kurchatov Institute'', Moscow, 117218 Russia}

\author{V.\,V.~Braguta}
\email[]{braguta@itep.ru}
\affiliation{Moscow Institute of Physics and Technology, Institutsky lane 9, Dolgoprudny, Moscow region, 141700 Russia}
\affiliation{Institute for Theoretical and Experimental Physics NRC ``Kurchatov Institute'', Moscow, 117218 Russia}
\affiliation{Bogoliubov Laboratory of Theoretical Physics, Joint Institute for Nuclear Research, Dubna, 141980 Russia}
\affiliation{Far Eastern Federal University, School of Biomedicine, 690950 Vladivostok, Russia}

\author{Massimo D'Elia}
\email{massimo.delia@unipi.it }
\affiliation{Dipartimento di Fisica dell'Universit\`a di Pisa, Largo B.~Pontecorvo 3, I-56127 Pisa, Italy}
\affiliation{INFN, Sezione di Pisa, Largo B.~Pontecorvo 3, I-56127 Pisa, Italy}
\author{A.\,Yu.~Kotov}
\email[]{andrey.kotov@phystech.edu}
\thanks{Corresponding author}
\affiliation{Moscow Institute of Physics and Technology, Institutsky lane 9, Dolgoprudny, Moscow region, 141700 Russia}
\affiliation{Bogoliubov Laboratory of Theoretical Physics, Joint Institute for Nuclear Research, Dubna, 141980 Russia}

\author{A.\,A.~Nikolaev}
\email[]{aleksandr.nikolaev@swansea.ac.uk}
\affiliation{Department of Physics, College of Science, Swansea University, Swansea SA2 8PP, United Kingdom}

\author{Francesco Sanfilippo}
\affiliation{INFN, Sezione di Roma Tre, Via della Vasca Navale 84, I-00146 Rome, Italy}
\email[]{francesco.sanfilippo@roma3.infn.it}

\begin{abstract}
 We study the electromagnetic (e.m.) conductivity of QGP in a magnetic background by lattice simulations with $N_f = 2+1$ dynamical rooted staggered fermions at the physical point. We study the correlation functions of the e.m.~currents at $T=200,\,250$\,MeV and use the Tikhonov approach to extract the conductivity. This is found to rise with the magnetic field in the direction parallel to it and to decrease in the transverse direction, giving evidence for both the Chiral Magnetic Effect and the magnetoresistance phenomenon in QGP. We also estimate the chiral charge relaxation time in QGP.
\end{abstract}
\maketitle

The Chiral Magnetic Effect (CME) is a well known anomaly-based phenomenon which can be realized in different systems with relativistic fermionic degrees of freedom~\cite{Fukushima:2008xe, Kharzeev:2013ffa, Kharzeev:2015znc}. The CME is the generation of a non-dissipative electric current along the external magnetic field in systems with a net imbalance between the number of right-handed and left-handed fermions or nonzero chiral density.\\
\indent The nonzero chiral density should be generated in order to experimentally observe the CME. In heavy-ion experiments the chiral density might be generated due to sphaleron transitions in the quark-gluon plasma (QGP)~\cite{Kharzeev:2007jp,Kotov:2018vyl}. In condensed matter systems the chiral density can be generated as the result of lattice deformations~\cite{Cortijo:2016wnf}. Another way to generate the chiral density and to observe the CME is to apply 
parallel electric and magnetic fields. In this case the chiral anomaly generates the imbalance between the right-handed and left-handed fermions which leads to the CME which manifests itself through the rise of electric conductivity along the magnetic field. This CME current has already been observed experimentally in condensed matter systems~\cite{Li:2014bha,Li2015,Li2016}.\\ \indent
Similarly to condensed matter systems, the latter mechanism can be realized in heavy-ion experiments, where colliding ions create hot QGP with deconfined relativistic quarks. In addition, in non-central collisions the QGP is affected by huge magnetic fields generated by the motion of colliding heavy ions~\cite{Kharzeev:2007jp}. As a result the electromagnetic (e.m.) conductivity of QGP along the magnetic field might be significantly enhanced. \\ \indent
Let us consider QGP in parallel electric $\bf E$ and magnetic $\bf B$ fields.  
Due to the axial anomaly these fields lead to the generation of a chiral density with the rate~\cite{Li:2014bha}
\beq
\frac {d \rho_5} {d t} = C \frac {e^2} {2 \pi^2} ~ {\bf E}\cdot {\bf B} - \frac {\rho_5} {\tau},
\label{rho5gen}
\eeq 
where $C=N_c\sum_{f}q^2_f$. 
The first term in Eq.~(\ref{rho5gen}) describes the production of chiral charge due to the chiral anomaly, while the second term stands for the decrease of chirality due to the chirality-changing processes with the relaxation time $\tau$. Note that Eq.~(\ref{rho5gen}) has the stationary solution
\beq
\rho_5 = C \frac {e^2} {2 \pi^2}  {\bf E}\cdot {\bf B} \tau,
\label{rho5}
\eeq
which describes the balance between anomaly based production rate and chirality relaxation processes.

The chiral charge density can be parameterized by the chiral chemical potential $\mu_5$ through the equation of state (EoS) $\rho_5=\rho_5(\mu_5)$. We use the linear response theory and consider the electric field $\bf E$ as a perturbation. In this limit the generated chiral chemical potential is small and the EoS reads
\beq
\rho_5=\mu_5 \chi(T, B)
+ O(\mu_5^3),
\label{eof}
\eeq  
where the $\chi(T, B)$ is a function of magnetic field and temperature. We mostly consider large magnetic fields ($q_f eB \gg T^2$), thus the chiral density is governed by the lowest Landau level degeneracy, $\chi \propto eB$ ($\chi = N_c \sum_f |q_f|\, eB/2\pi^2$ in the non-interacting approximation). The CME generates the electric current
\beq
{\bf j_{CME}}=C \frac {e^2} {2 \pi^2} \mu_5 {\bf B}.
\label{cme}
\eeq
Combining Eq.~(\ref{rho5}), Eq.~(\ref{eof}) and Eq.~(\ref{cme}) one obtains the conductivity due to the CME
\beq
j_{CME}^i =  \sigma_{CME}^{ij} E^{j},~ \sigma_{CME}^{ij}= C^2 \frac {e^4} {4 \pi^4} \frac {\tau} {\chi(T,B)} B^{i} B^{k}.
\label{cme_cond}
\eeq 
It is assumed that the magnetic field is applied along the $z$ axis. 

In addition to the CME current there is also Ohmic current in the system. The total conductivity is the sum of Ohmic and the CME conductivities $\sigma=\sigma^{O}+\sigma^{CME}$. If the electric field is applied along the $x$ axis, the Lorentz force reduces the transverse conductivity $\sigma^{O}_{xx}$. The $\sigma^{CME}_{xx}$ component is zero in this case. The decrease of $\sigma_{xx}$ in external magnetic field is called magnetoresistance. 
On the other hand, if electric field is applied along the magnetic field, there is no Lorentz force and magnetoresistance. 

At the same time $\sigma^{CME}_{zz}$ is a rising function of the magnetic field which can be a manifestation of the CME\footnote{In what follows the transverse conductivity $\sigma_{xx}$ will be designated as $\sigma_{\perp}$, while the conductivity along magnetic field $\sigma_{zz}$ will be designated as $\sigma_{\parallel}$}.
These facts allow one to expect that the transport properties of QGP in heavy ion collision experiments can be considerably modified by the external magnetic field. Since the transport properties of QGP are particularly important for understanding of heavy ion collision phenomenology, in this paper we are going to study the conductivity of QGP in external magnetic field.

It should be noted that the e.m.~conductivity of QCD was calculated in a number of lattice studies (see for instance~\cite{Amato:2013naa, Aarts:2014nba, Brandt:2015aqk, Ding:2016hua}). At the same time, some e.m.~properties of the QGP in the presence of a magnetic background, like its magnetic susceptibility,
have been already explored~\cite{Bonati:2013lca,Levkova:2013qda,Bonati:2013vba,Bali:2013owa,Bali:2014kia,Hattori:2016cnt,Fukushima:2019ugr,Braguta:2019yci}, as well as the emergence of anisotropies related to the magnetic background in other relevant quantities~\cite{Bali:2013owa,Bonati:2014ksa,Bonati:2016kxj,Bonati:2017uvz,Bonati:2018uwh}.
Quenched lattice study of the e.m.~conductivity of QCD in external magnetic field was carried out in~\cite{Buividovich:2010tn}, where no sign of neither CME nor magnetoresistance in QGP was found. We would like also to mention the lattice study of the e.m.~conductivity with the external magnetic field in the Dirac semimetals~\cite{Boyda:2017dml} where the CME and magnetoresistance were observed in the semimetal phase which is similar to QGP in some properties. Finally we would like to mention lattice study of the CME in thermodynamic equilibrium~\cite{Yamamoto:2011gk}.

In this paper we carry out the first lattice study of the e.m.~conductivity of QGP in external magnetic field with $N_f = 2+1$ dynamical staggered quarks at physical quark masses, more details about the lattice discretization and algorithms are provided in Appendix. We  consider temperatures $T=200,\,250\,$MeV for several values of the external magnetic field. Most simulations are carried out on a $16\times64^3$ lattice, with spacings $a=0.0618$\,fm and $a=0.0493$\,fm correspondingly. To check lattice spacing dependence we also consider a $10\times48^3$ lattice with $a=0.0988$\,fm. To study the ultraviolet (UV) properties of the correlator of two e.m.~currents we consider simulations on a $96\times48^3$ lattice at $a=0.0988$\,fm, which corresponds to approximately zero temperature.

To study the conductivity we apply the following strategy. We first calculate the lattice correlation function 
\beq
C_{ij}(\tau) = \frac {1} {L_s^3} \langle J_i(\tau) J_j(0) \rangle,
\label{stag}
\eeq
where $\tau$ is the Euclidean time and $J_i(\tau)$ is the conserved current
\beq
J_{i}(\tau)= \frac 1 4 e \sum_f q_f \sum_{\vec x} \eta_{i}(x) \bigl ( \bar \chi^f_{x} U_{x,i} \chi^f_{x+i} + \bar \chi^f_{x+i} U_{x,i}^\dagger \chi^f_{x} \bigr ),
\label{eq:current}
\eeq
where $x=(\tau, \vec x)$, $\eta_i(x)=(-1)^{x_1+..x_{i-1}}$, $i=1,2,3$, $\bar \chi^f_{x}, \chi^f_{x}$ are staggered fermion fields of $f=u,\,d,\,s$ flavours, and $U_{x,i}$ is the gauge field matrix.

The well known property of the staggered fermions is that the correlator~(\ref{stag}) corresponds to two different operators for the even $\tau=2n \times a$ and odd $\tau=(2n+1) \times a$ slices. In the continuum limit $C_{ij}(\tau)$ reads 

\beq
\label{correlator}
C^{\mbox{\footnotesize{e}},\,\mbox{\footnotesize{o}}}_{ij}(\tau)=\sum_{\vec x} \left(\langle A_i(x) A_j(0) \rangle - s^{\mbox{\footnotesize{e}},\,\mbox{\footnotesize{o}}} \langle B_i(x) B_j(0) \rangle \right),
\eeq
where $s^{\mbox{\footnotesize{e}},\,\mbox{\footnotesize{o}}} = \pm 1$ is the timeslice parity and 
\beq
\nonumber
A_i= e \sum_f q_f \bar \psi^f \gamma_i \psi^f, \quad B_i = e \sum_f q_f \bar \psi^f \gamma_5 \gamma_4 \gamma_i  \psi^f,
\eeq
and $\psi^f$ is Dirac spinor of the flavour $f$.  Notice that the operator $A_i$ corresponds to e.m.~current in the continuum whereas we would like to remove the $B_i$ contribution. 

Next let us recall that the current-current Euclidean correlators both for even and odd slices $C^{\mbox{\footnotesize{e}},\,\mbox{\footnotesize{o}}}_{ij}$ are related to its spectral functions $\rho^{\mbox{\footnotesize{e}},\,\mbox{\footnotesize{o}}}_{ij}(\omega)$ as 
\begin{equation}
	C^{\mbox{\footnotesize{e}},\,\mbox{\footnotesize{o}}}_{ij}(\tau)=\int_0^{\infty} \frac{d\omega}{\pi} K(\tau, \omega) \rho^{\mbox{\footnotesize{e}},\,\mbox{\footnotesize{o}}}_{ij}(\omega),
	\label{eq:Kubo}
\end{equation}
where $K(\tau, \omega)=\frac{\cosh\omega(\beta-\tau/2)}{\sinh\omega\beta/2}$. The e.m.~conductivity $\sigma_{ij}$ is related to the spectral densities $\rho^{\mbox{\footnotesize{e}},\,\mbox{\footnotesize{o}}}_{ij}(\omega)$ through the Kubo formulas
\begin{equation}
	\frac{\sigma_{ij}}{T} = \frac{1}{2 T} \lim\limits_{\omega \to 0} \frac 1 {\omega} \biggl ( {\rho^{\mbox{\footnotesize{e}}}_{ij}(\omega)} + {\rho^{\mbox{\footnotesize{o}}}_{ij}(\omega)} \biggr ). 
	\label{eq:limit}
\end{equation}

Notice that in last formula the contribution of the correlator $\langle B_i(\tau) B_j(0) \rangle$ to the sum $\rho^{e}_{ij}+\rho^{o}_{ij}$ cancels out and in the continuum limit the e.m.~conductivity is reproduced.  It is important to notice that similarly to~\cite{Amato:2013naa, Aarts:2014nba, Brandt:2015aqk, Ding:2016hua, Buividovich:2010tn} in this calculation of the correlation function (\ref{stag}) 
only connected diagrams are accounted. 

Given the correlation functions $C_{ij}^{\mbox{\footnotesize{e}},\,\mbox{\footnotesize{o}}}$ one needs to invert the integral equation~(\ref{eq:Kubo}) and determine the spectral functions $\rho_{ij}^{\mbox{\footnotesize{e}},\,\mbox{\footnotesize{o}}}$ to find the conductivity. To do this one can apply the  Backus-Gilbert (BG)~\cite{Backus:1} or Tikhonov regularization (TR)~\cite{Tikhonov:1963} approaches. A detailed description of these approaches can be found in Appendix. Our calculation shows that both approaches give similar results but the TR resolution function for the conductivity is a little narrower (see below). For this reason we calculate the conductivity within the TR approach. 

The TR method is based on the calculation of the estimator of the spectral function $\tilde \rho( \bar \omega)$ instead of the spectral function $\rho(\omega)$ itself. The estimator is defined as
\begin{equation}
   \biggl (	\frac {\tilde \rho_{ij}(\omega)}{\omega} \biggr )_{\omega=\bar \omega} = \int d \omega \delta(\bar \omega, \omega) \frac {\rho_{ij}(\omega)} {\omega},
	\label{eq:contraction}
\end{equation}
where $\delta(\omega, \bar \omega)$ is the resolution function peaked around $\bar \omega$. 
If $\delta(\omega, \bar \omega) = \delta ( \omega - \bar \omega)$ the estimator of the spectral function exactly reproduces the spectral function $\tilde \rho(\bar \omega) = \rho(\bar \omega)$. However, in real calculations the resolution function has a finite width of few $T$.  In particular, at the $\bar \omega = 0$ the width of the resolution  function is $\sim 3.5\,T$. The estimator averages the spectral function over the width of the resolution function.

\begin{figure}[h!]
    \centering
    \includegraphics[scale=0.45]{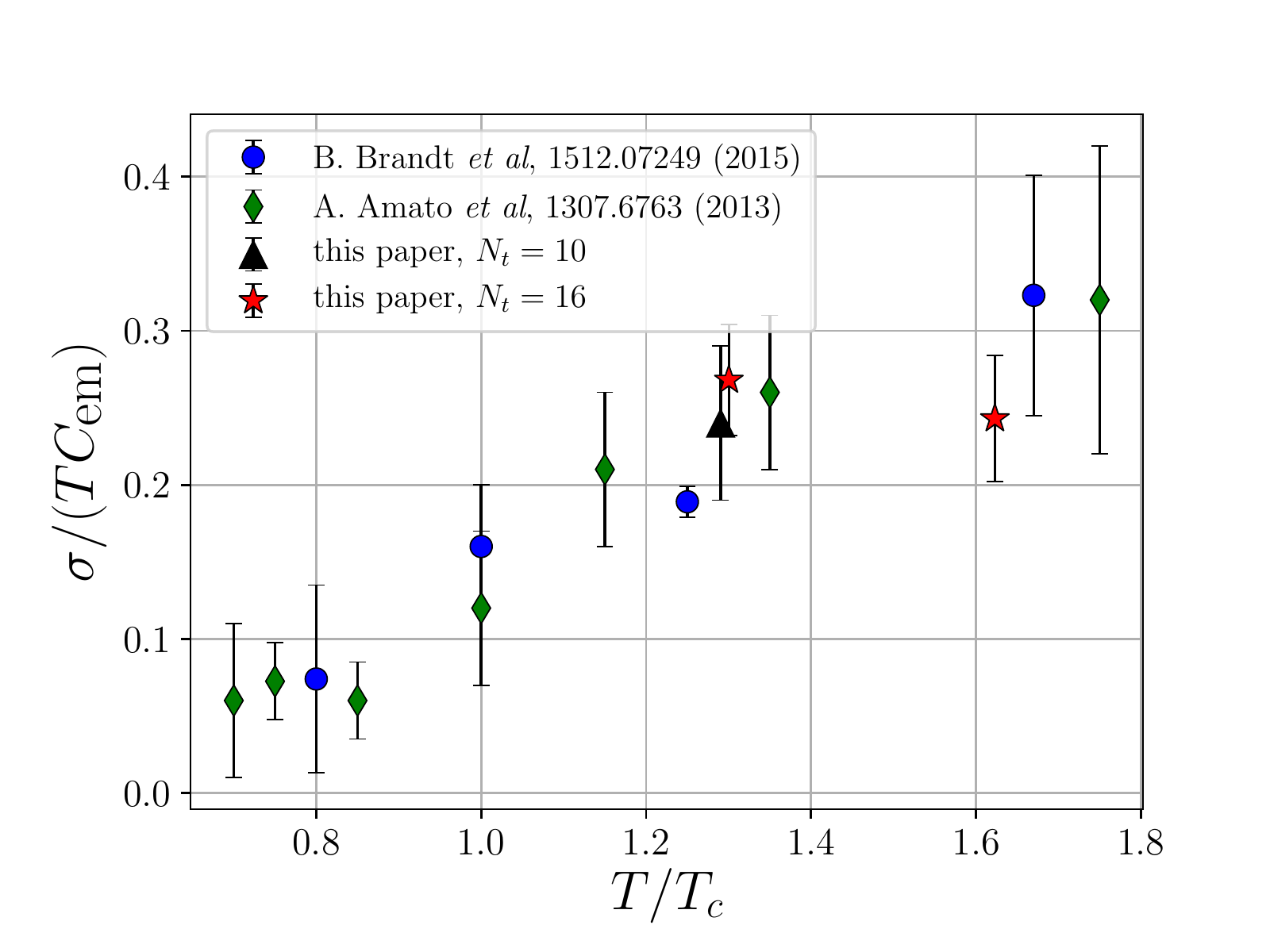}
    \caption{The e.m. conductivity in QCD as a function of temperature normalized to the factor $T C_{em}$ at $eB = 0$. The green rhombi show the $n_f = 2 + 1$ data from~\cite{Amato:2013naa}, the blue circles stand for the $n_f = 2$ data from~\cite{Brandt:2015aqk}. The red stars show the results of this paper calculated at temperatures $T = 200,\,250\,\mbox{MeV}$ on the lattice $16\times64^3$. The black triangle shows the result for $T = 200\,\mbox{MeV}$ calculated on the lattice $10\times48^3$.}
    \label{fig:compare_B0}
\end{figure}

The TR method can be used to reconstruct the spectral function at $\omega = 0$ if the width of the resolution function $\delta(\bar \omega=0, \omega)$ is of the order of or smaller than the characteristic variation scale of the spectral function around $\omega = 0$,  otherwise the TR method might underestimate it. Lattice data for the correlation functions of the e.m. currents are well described by either the anzats combining the transport peak at small frequencies and UV contribution at large frequencies~\cite{Aarts:2014nba, Brandt:2015aqk,Ding:2016hua} or by the AdS/CFT spectral function~\cite{Ding:2016hua}.
In the temperature interval under consideration the 
widths of the resolution functions are close or smaller than the variation scale of the spectral functions from~\cite{Aarts:2014nba, Brandt:2015aqk,Ding:2016hua}. For this reason we believe that the TR method can be used to calculate the e.m.~conductivity in QGP. Notice also that our results for the conductivity at zero magnetic field shown in Fig.~\ref{fig:compare_B0} are in agreement with the previous studies which hints  at a correct reconstruction of the
conductivity.

Another important issue is the UV contribution to the reconstructed conductivity. For instance, in the studies of shear and bulk viscosities of gluon plasma~\cite{Astrakhantsev:2017nrs, Astrakhantsev:2018oue} the UV spectral density scales as $\rho \propto \omega^4$, which results in a large UV contribution to the estimator~(\ref{eq:contraction}). This contribution should be subtracted in order to obtain reliable results. For the conductivity the UV contribution scales as $\rho \propto \omega^2$ and our calculation shows that the UV gives $\sim 20-30\,\%$ contribution at $\bar \omega=0$. In Appendix we give a detailed description of the UV subtraction procedure. 

To summarize, the calculation is done in the following steps. First we measure lattice correlation functions $C^{\mbox{\footnotesize{e}},\,\mbox{\footnotesize{o}}}_{ij}(\tau)$. Then we calculate the estimators ${\tilde \rho^{\mbox{\footnotesize{e}},\,\mbox{\footnotesize{o}}}(\bar \omega)}/{\bar \omega}$ at $\bar \omega =0$ within the TR approach and subtract the UV contribution. Finally, using Eq.~(\ref{eq:limit}), we calculate the e.m.~conductivity. 

The e.m.~conductivities normalized to the factor $T C_{em}$ ($C_{em}=e^2\sum_f q_f^2$) at zero magnetic field and temperatures $T=200,\,250$\,MeV are shown in Fig.~\ref{fig:compare_B0}. In addition we plot the results of~\cite{Amato:2013naa,Brandt:2015aqk}. Notably our results are in agreement with previous lattice studies within the uncertainties. 

Let us now consider the e.m.~conductivity of QGP in the presence of the external magnetic field. From a technical point of view, the magnetic field affects
directly the path-integral measure and the fermion propagators entering the construction of the e.m.~currents; moreover, the e.\,m.~$U(1)$ phases enter the gauge links in the definition of the split current in Eq.~(\ref{eq:current}). Apart from this, the problem turns out to be easier than at $eB = 0$. In particular, instead of the correlation functions $C^{\mbox{\footnotesize{e}},\,\mbox{\footnotesize{o}}}_{eB}$ we consider the difference $\Delta C^{\mbox{\footnotesize{e}},\,\mbox{\footnotesize{o}}} = C^{\mbox{\footnotesize{e}},\,\mbox{\footnotesize{o}}}_{eB} - C^{\mbox{\footnotesize{e}},\,\mbox{\footnotesize{o}}}_{eB=0}$. Since, for the chosen values of the lattice spacing, the UV regime starts at $\omega_0 \sim 2\,$GeV, we note that $q_f eB \ll \omega^2$ for all frequencies in the UV regime and magnetic fields. Thus, one can consider the UV spectral function magnetic field-independent and assume that the differences $\Delta C^{\mbox{\footnotesize{e}},\,\mbox{\footnotesize{o}}}$ do not contain the UV contribution. The results for $\Delta C^{\mbox{\footnotesize{e}},\,\mbox{\footnotesize{o}}}$ turn out to be more accurate since the UV--estimation uncertainty is absent in this case. The correlator $\Delta C^{\mbox{\footnotesize{e}},\,\mbox{\footnotesize{o}}}$ is related to additional conductivity due to the presence of the magnetic field. In our further study we apply the TR approach to the differences $\Delta C^{\mbox{\footnotesize{e}},\,\mbox{\footnotesize{o}}}$.

\begin{figure}[h!]
    \centering
    \includegraphics[scale=0.45]{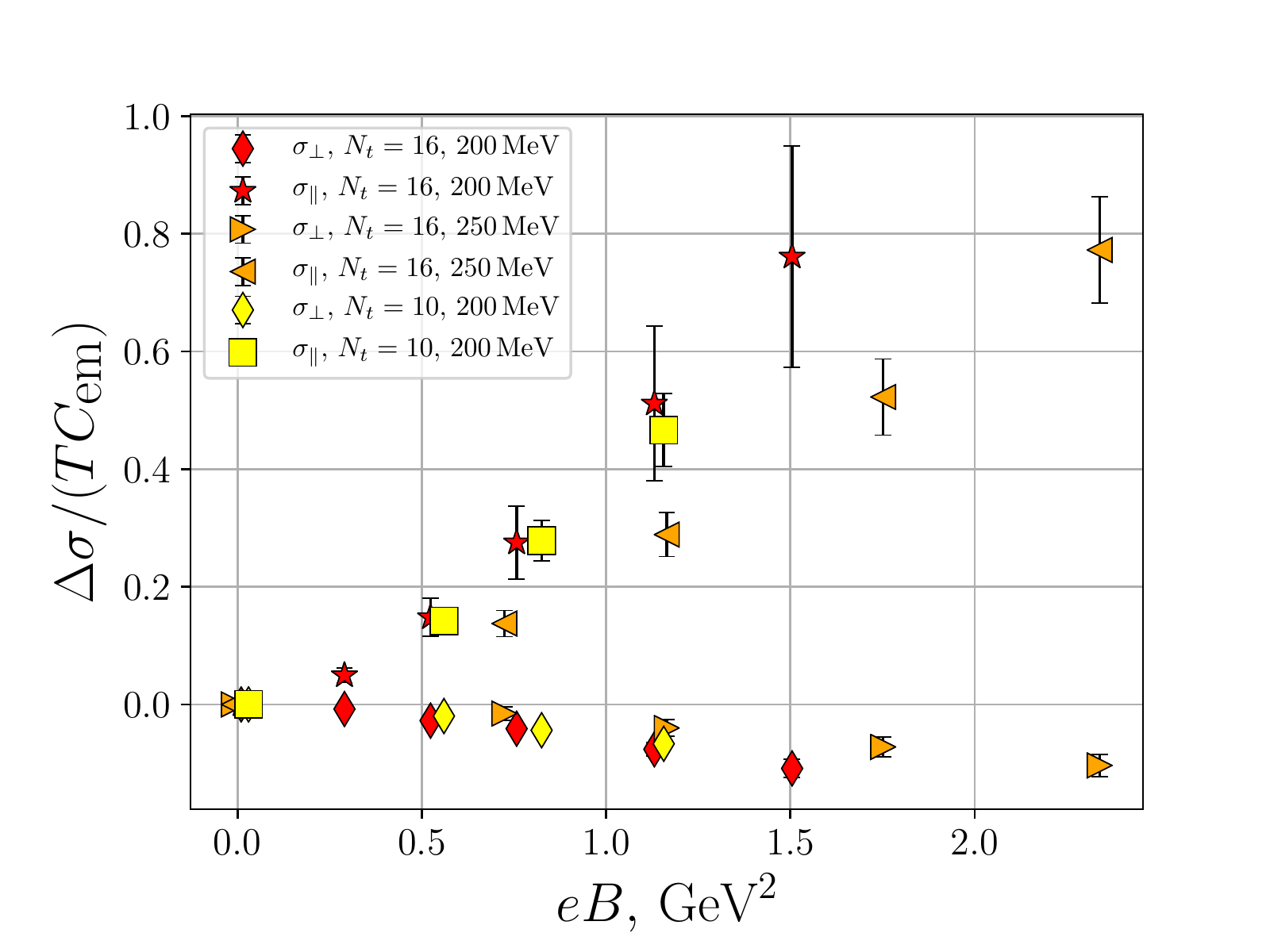}
    \caption{The e.m. conductivities due to the external magnetic field $\Delta \sigma= \sigma_{eB}-\sigma_{eB=0}$ normalized to $T C_{em}$ at temperatures $T=200,\,250$\,MeV. The $\Delta \sigma_{\parallel}, \Delta \sigma_{\perp}$ correspond to the parallel and transverse directions to the magnetic field. We also show the result for $T = 200\,$MeV calculated on the lattice with $N_t=10$ to check the finite $N_t$ artifacts.
    }
    \label{fig:CME}
\end{figure}

The e.m. conductivity due to the external magnetic field $\Delta \sigma= \sigma_{eB}-\sigma_{eB=0}$ normalized to $TC_{em}$ at temperatures $T=200,\,250$\,MeV  is shown in Fig.~\ref{fig:CME}. It is seen that $\Delta \sigma_{\parallel}$ rises with magnetic field for both temperatures, which is the observation of the CME in QGP on the lattice. {Notice also that the rise of the $\Delta \sigma_{\parallel}$ becomes linear for sufficiently large magnetic field,  in agreement with Eq.~(\ref{cme_cond})}. A similar linear growth of parallel conductivity $\sigma_{\parallel}$ with large magnetic fields was obtained in~\cite{Fukushima:2017lvb} by studying kinetic equations. In turn, $\Delta \sigma_{\perp}$ is negative and decreases with the magnetic field, which is the observation of the magnetoresistance in QGP. Note also that the slopes on both functions $\Delta \sigma_{\parallel}(eB), \Delta \sigma_{\perp}(eB)$ decrease with temperature. {We believe that this can be explained by a decrease of the relaxation time with temperature because of the increased thermal activity. }

In order to estimate the finite $N_t$ effects we calculate the conductivity at $T=200$\,MeV, $eB=0,\,0.52,\,0.79,\,1.12\,\mbox{GeV}^2$ on the lattice $10\times48^3$ in addition to the lattice $16\times64^3$ at hand. The results of this calculation are shown in Fig.~\ref{fig:compare_B0} and Fig.~\ref{fig:CME}. It is seen that within the uncertainties of the calculation the conductivities calculated at different lattice spacings are in agreement with each other.  From this we conclude that discretization effects are under control in our study.

\begin{figure}[h!]
    \centering
    \includegraphics[scale=0.45]{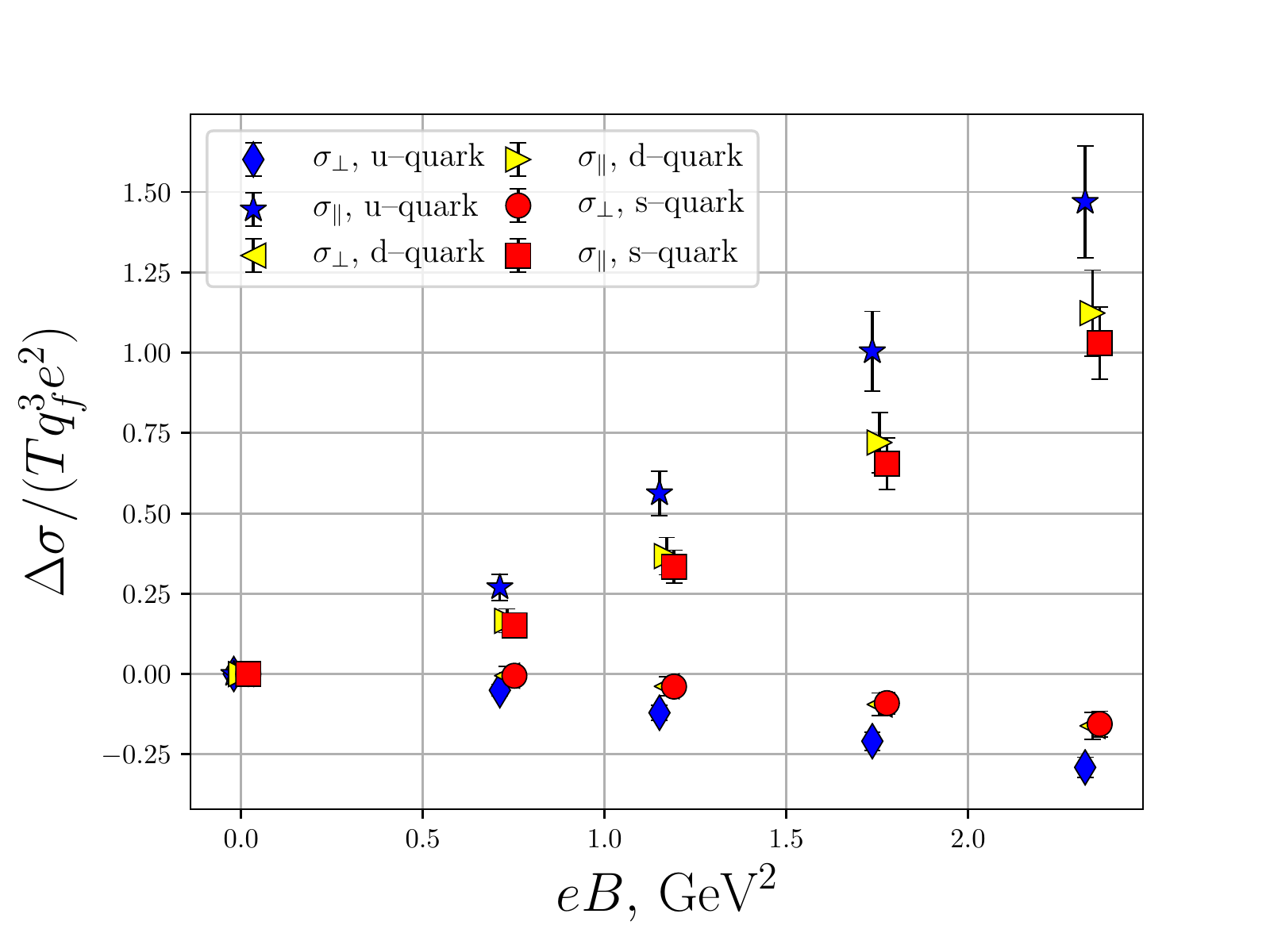}
    \caption{The u--, d-- and s--quark contributions to the conductivities $\Delta \sigma_{\parallel}, \Delta \sigma_{\perp}$ normalized to the factor $T e^2 q_f^3$ at $T=250$\,MeV.}
    \label{fig:conductivity_uds}
\end{figure}

From equations~(\ref{stag}),~(\ref{eq:current}) it is seen that there are two contributions to the conductivity. The first one is the valence quarks contribution to the current operator~(\ref{eq:current}) and the other results from the sea quarks in the fermionic determinant of the partition function. The valence quarks' contribution can be separated into u--, d-- and s--quark contributions to the conductivity which can be calculated from the quark loop of the corresponding flavour\footnote{
Notice that the separation of the conductivity to 
into each flavour contribution is possible only for the connected diagrams.}. In Fig.~\ref{fig:conductivity_uds} we plot the u--, d-- and s--quark contributions to the conductivities $\Delta \sigma_{\parallel}, \Delta \sigma_{\perp}$ normalized to the factor $T e^2 q_f^3$ at temperature $T=250$\,MeV. The normalization factor was chosen so as to reduce the dependence of the corresponding contribution on the quark flavour: the $q_f^2$ results from the correlation function~(\ref{stag}) while the additional $q_f$ results from the leading order coupling of the magnetic field to the quark $q_f e H$. From Fig.~\ref{fig:conductivity_uds} it is seen that within the uncertainty we do not see the dependence of  $\Delta \sigma_{\perp}/q_f^3$ on the quark flavour. In turn, within the uncertainty the contributions of the d-- and s-- quarks to $\Delta \sigma_{\parallel}/q_f^3$ agree, while the contribution of the u--quark is slightly larger. This can be explained by the larger charge of the u-quark. We thus conclude that the leading dependence of $\Delta \sigma_{\parallel}, \Delta \sigma_{\perp}$ on the quark flavour is proportional to $q_f^3$. In addition the relatively heavy s--quark mass does not influence $\Delta \sigma_{\parallel}, \Delta \sigma_{\perp}$ within the uncertainties. 

\begin{figure}[h!]
    \centering
    \includegraphics[scale=0.45]{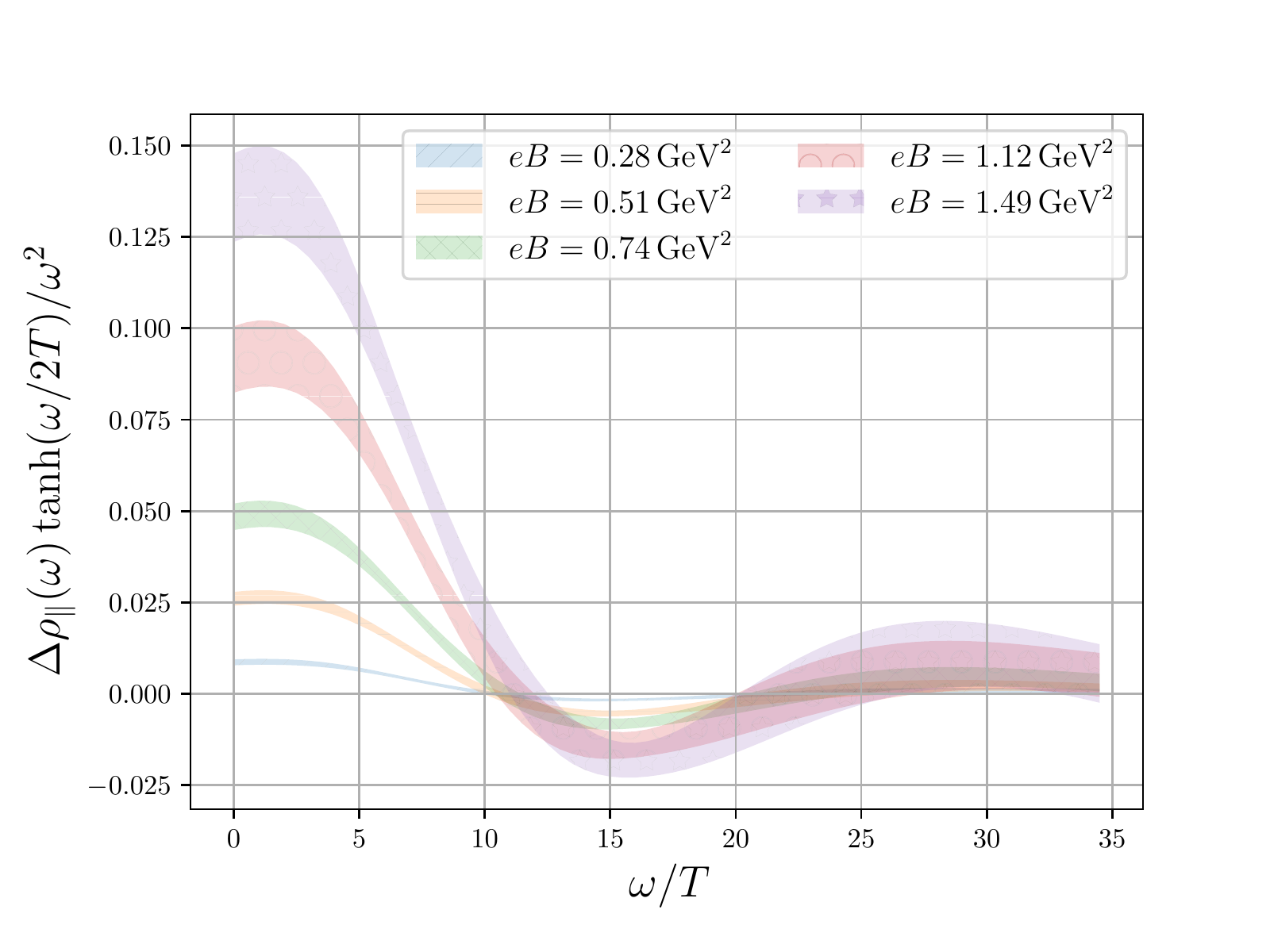}
    \caption{The reconstructed spectral function $\Delta \rho_{\parallel}$ from the difference of the correlation functions $\Delta C^{\mbox{e},\,\mbox{o}}$ at temperature $T=200$\,MeV for various magnetic fields.}
    \label{fig:peak_at_zero}
\end{figure}

The TR method also allows to reconstruct the spectral function, for instance $\Delta \rho_{\parallel}$. The reconstructed $\Delta \rho_{\parallel}(\omega)/f(\omega)$, where $f(\omega) = \omega^2/\tanh{\omega/2T}$, at $T=200$\,MeV is shown in Fig.~\ref{fig:peak_at_zero}. It is seen that the infrared part of the spectral function ($\omega/T < 10$) has a transport peak with the height rising with the magnetic field. The ultraviolet part of the spectral function weakly depends on the magnetic field and remains close to zero as expected.  

The dependence of $\Delta \sigma_{\parallel}$ on the magnetic field which is responsible for the CME allows us to estimate the relaxation time of the chiral charge (see Eq.~(\ref{cme_cond})). The relaxation times turn out to be $\tau(200\,\mbox{MeV})=0.26(5)\,\mbox{fm/c}$, $\tau(250\,\mbox{MeV})=0.24(3)\,\mbox{fm/c}$,
which is in agreement with the relaxation time obtained in~\cite{PhysRevD.94.054011}, where $\tau$ lies in the interval $\sim 0.1-1.0\,\mbox{fm/c}$ at $T \sim 200-250\,$MeV depending on the model. 

The e.m.~conductivity of QCD in external magnetic field was studied in~\cite{Buividovich:2010tn} in the quenched approximation, reporting no evidence of either CME or magnetoresistance in QGP. A possible source of the disagreement is the small magnetic field used in~\cite{Buividovich:2010tn}, where 
the largest field used in the deconfinement phase is $eB = 0.36$\,GeV$^2$: at the same value our signal is quite small (see Fig.~\ref{fig:CME}), so probably the signal was hardly detectable in~\cite{Buividovich:2010tn}. The authors of~\cite{Buividovich:2010tn} have also conducted simulations in the confinement phase and observed the rise of $\Delta \sigma_{\parallel}$ and drop of the $\Delta \sigma_{\perp}$. Similarly, we have also calculated the conductivities in the confinement phase using the approach developed in this paper and obtained similar results: $\Delta \sigma_{\parallel}$ rises while $\Delta \sigma_{\perp}$ drops with magnetic field. 
However, we would like to stress that contrary to the deconfinement phase the structure of the spectral function in the confinement phase is rather complicated. For instance, it contains the contribution of the intermediate $\pi^+\pi^-$ mesons or the $\rho$ meson peak which has a large spectral weight in the confinement~\cite{Brandt:2015aqk}. It is reasonable to expect that the external magnetic field modifies the spectral function, for instance through the light meson masses modification~\cite{Bali:2017ian, Luschevskaya:2016epp, Andreichikov:2016ayj}. Thus, in order to check the presence of CME in the confinement phase one has to separate the contribution to the spectral function due to the conductivity $\omega \sim 0$ from the contribution of the light mesons $\omega \sim 2 m_{\pi}, m_{\rho},...$. This is a difficult task which can not be done with the data used in this paper;  that might be the case for Ref.~\cite{Buividovich:2010tn} as well. Note also that in~\cite{Boyda:2017dml} $\Delta \sigma_{\parallel}$ was studied in Dirac semimetals both in semimetal and insulator phases. Due to chiral symmetry breaking in the insulator phase it was found that $\Delta \sigma_{\parallel}=0$ and there is no CME in this phase. In the confinement phase there is also chiral symmetry breaking. For this reason one can expect that there is no CME in the confinement phase at sufficiently low temperature. 

{ In conclusion, this paper is devoted to a lattice study of the e.m.~conductivity of QGP in the presence of a magnetic background field.}
{ It is found that the conductivity along the magnetic field rises with the magnetic field, which is a possible manifestation of the CME. On the contrary, the conductivity in the transverse direction is decreasing with the magnetic field, which is the magnetoresistance phenomenon. Thus we observe evidence for the CME and magnetoresistance in QGP. Finally, we have also computed the relaxation time of the chiral charge in QGP 
for the explored temperature range.}

\section{ACKNOWLEDGMENTS}

The work of N.\,Yu.\,A., V.\,V.\,B. and A.\,Yu.\,K. was supported by RFBR grants 18-02-01107 and 18-02-40126. A.\,A.\,N. acknowledges the support from STFC via grant ST/P00055X/1. {  Numerical simulations have been carried out on the MARCONI machine at CINECA, based on the agreement between INFN and CINECA (under project INF19\_npqcd), and on the computing resources} of the federal collective usage center Complex for Simulation and Data Processing for Mega-science Facilities at NRC ``Kurchatov Institute'',~\url{http://ckp.nrcki.ru/}. In addition, the authors used  the supercomputer of Joint Institute for Nuclear Research ``Govorun''.

\appendix
\subsection{Appendix A: Description of the Backus-Gilbert and Tikhonov regularization methods}
\label{appendixA}

Maximal Entropy Method (MEM) is a popular method for the reconstruction of the spectral functions~\cite{Asakawa:2000tr}. For the e.m. conductivity MEM was applied in papers~\cite{Buividovich:2010tn, Aarts:2014nba}. It is rather difficult to carry out our study with MEM. This is because for staggered fermions we have $N_t/2=8$ points(due to the periodicity of the correlator)
in temporal direction which are splitted into 4 points for even time slices and 4 points for odd time slices.
To conduct the reconstruction in this case you have to reconstruct separately even and odd spectral functions. We believe that for MEM this is very complicated task.
Notice also that MEM can be applied only for positive spectral functions. However, this is not the case for odd branch of the spectral function in magnetic field. For these reasons we decided to apply Backus-Gilbert(BG) and Tikhonov regularization (TR) methods.

The BG and TR methods are non-parametric approaches which can be used to study the spectral function\footnote{ The BG and TR methods were used to study transport properties of different strongly correlated systems in~\cite{Brandt:2015aqk, Astrakhantsev:2017nrs, Astrakhantsev:2018oue, boyda2016many}}. These methods are aimed at the solution of the equation $$C(\tau) = \int\limits_{0}^{+\infty}\frac{d\omega}{2 \pi} \frac{\rho(\omega)}{f(\omega)} K(\omega, \tau), $$
where $ K(\omega, \tau) = \frac{\cosh \omega \left(\tau - \beta / 2\right)}{\sinh\omega \beta / 2} f(\omega)$ and $f(\omega)$ is an arbitrary function. 
Within the BG and TR methods instead of $\rho(\omega)$ one reconstructs the estimator $\bar \rho(\bar \omega)$ expressed as 
\beq
\bar \rho(\bar \omega) = f(\bar \omega) \int\limits_0^{\infty} d \omega \delta (\bar \omega, \omega) \frac {\rho(\omega)} {f(\omega)}, 
\label{barf}
\eeq
where $\delta(\bar \omega, \omega)$ is the resolution function that  has a peak around $\bar \omega$ and normalized. The BG and TR are the linear methods and the resolution function is taken in the form
\beq
\delta(\bar \omega, \omega) = \sum_i q_i(\bar \omega) K(x_i, \omega),
\eeq
thus the estimator is a linear combination of the correlation function values
\beq
\bar \rho(\bar \omega) = f(\bar \omega) \sum_i q_i(\bar \omega) C(\tau_i).
\label{barrho} 
\eeq
Accurate reconstruction of $\rho(\omega)$ requires minimization of the width of $\delta(\bar \omega, \omega)$. However, too small values of the estimator make the method unstable and susceptible to noise in the data. Thus, the method requires regularization that should be properly adjusted.

Within the BG method one minimizes the functional $\displaystyle \mathcal{H}(\rho(\omega)) = \lambda \mathcal{A}(\rho(\omega)) + (1 - \lambda) \mathcal{B}(\rho(\omega))$. The term $\mathcal{A}$ represents the width of the resolution function: $\mathcal{A} = \int_0^{\infty} d \omega \delta(\bar \omega, \omega) (\omega - \bar \omega)^2$. The term $\mathcal{B}(\rho(\omega)) = \mbox{Var}[\rho(\omega)]$ regularizes $\rho(\omega)$ making it less susceptible to noise. In terms of the covariance matrix and functions $q_i(\bar \omega)$ used to define $\bar \rho(\bar \omega)$ in Eq.~(\ref{barrho}), it reads $\displaystyle \mathcal{B}(\vec{q}) = \vec{q}^T \hat{S} \vec{q}$. Thus, statistical uncertainties are reduced at cost of increasing the width of the resolution function through decrease of $\lambda$. 

The minimization of $\mathcal{H}$ gives the following linear functions on the form (\ref{barrho}) 
\begin{gather}
q_i(\omega)  =  \frac { \sum_j W^{-1}_{ij} (\bar \omega) R(x_j) } { \sum_{kj} R(x_k) W^{-1}_{kj} (\bar \omega) R(x_j) }, \\
W_{ij}(\bar \omega)  =  \lambda\int\limits_{0}^{\infty} d \omega K(x_i,\omega) (\omega- \bar \omega)^2 K(x_j,\omega) + (1 - \lambda) S_{i j}, \\
R(x_i)  =  \int\limits_0^{\infty} d \omega K(x_i,\omega). 
\label{eq:BG_formulas}
\end{gather}

The TR method is another way of the regularization of the same problem. While in the BG method the regularization is performed as $W_{ij} \to \lambda S_{ij} + (1 - \lambda) W_{ij}$, in the TR scheme the SVD decomposition of $W^{-1} = V D U^{T}$ is regularized. The diagonal matrix $D = \mbox{diag}\left(\sigma_1^{-1}, \sigma_2^{-1}, \ldots, \sigma_n^{-1}\right)$ might have very large entries that represent the susceptibility of the data to noise. The regularization is done by adding the regularizer $\gamma$ to all entries as $\tilde D = \mbox{diag}\left((\sigma_1 + \gamma)^{-1}, (\sigma_2 + \gamma)^{-1}, \ldots, (\sigma_n + \gamma)^{-1}\right)$. Thus, small $\sigma_i$ will be smoothly cut-off.

\begin{figure}[h!]
    \centering
    \includegraphics[scale=0.5]{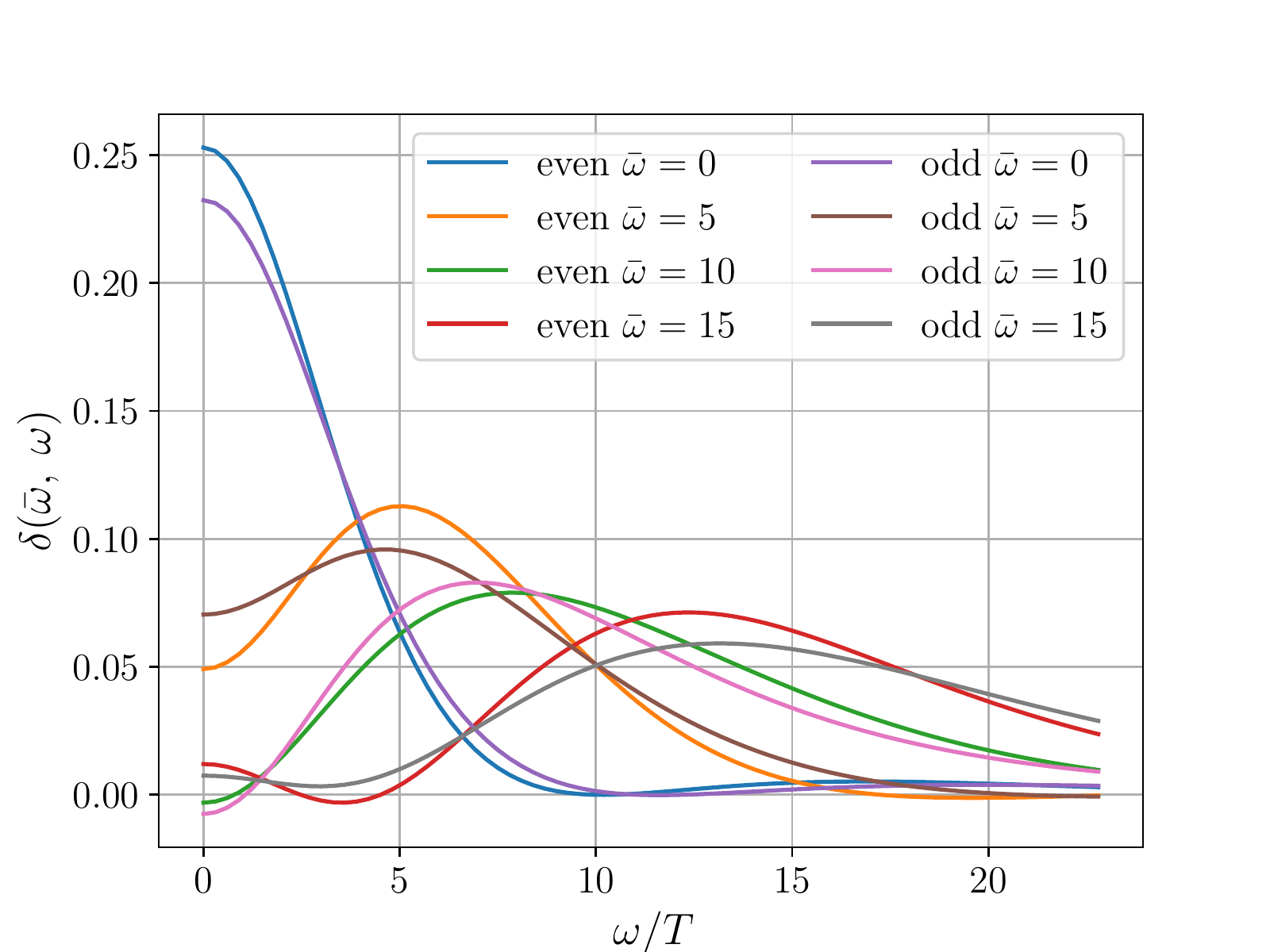}\\\includegraphics[scale=0.5]{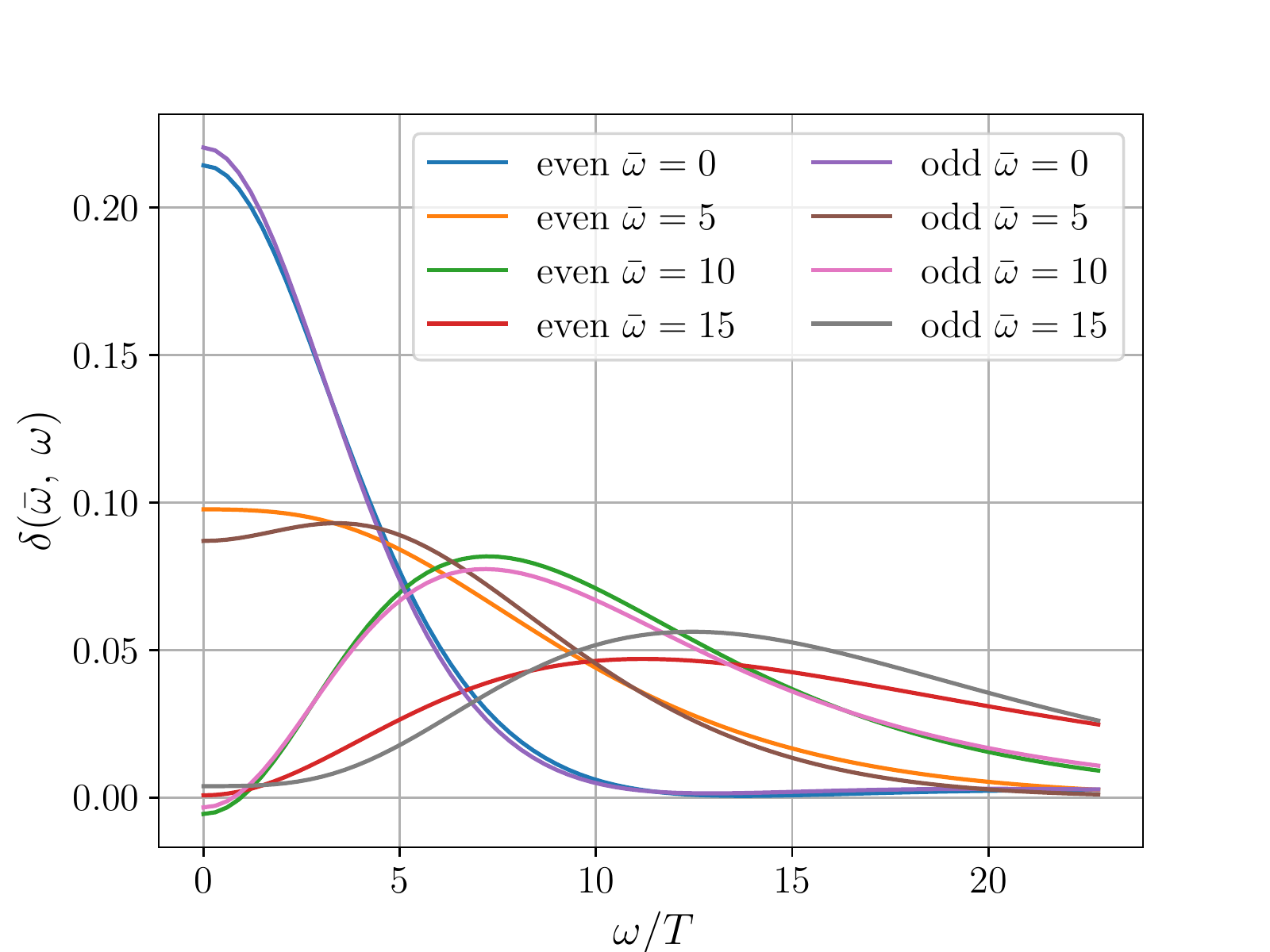}
    \caption{(Upper): the resolution functions for the BG regularization at $T = 200\,\mbox{MeV}$ and $\lambda = 0.01$.
    (Bottom): the resolution functions for the TR regularization at $T = 200\,\mbox{MeV}$ and $\gamma = 1.0$.}
    \label{fig:deltas}
\end{figure}

In Fig.~\ref{fig:deltas} we plot typical resolution functions for BG and TR regularizations at different $\bar \omega$ for the $\lambda = 0.01$ and $\gamma = 1$.

We remark that both in the BG and TR methods, the resolution function is an outcome of the method itself, and cannot be chosen a priori. This makes it difficult to well define a continuum limit, since there is no guarantee that the measured quantity is defined in the same way across different gauge ensemble. However, in our calculation we empirically observe that the dependence of the resolution functions on the parameters of the calculation is very weak. This is reflected in the good agreement of the results of $N_t=10$ and $N_t=16$ lattices. In the future, if a continuum limit has to be carried out, a fixed resolution function
must be employed, for example following the approach suggested in~\cite{Hansen:2019idp}.

\appendix
\subsection{Appendix B: Reconstruction of the spectral function}
\label{appendix:appendix_uv}

As mentioned, the calculation of the electromagnetic conductivity is carried out in the following steps. Firstly, we measure the lattice correlation functions $C^{\mbox{\footnotesize{e}},\,\mbox{\footnotesize{o}}}_{ij}(\tau)$ (\ref{correlator}). 
Then we calculate the estimators ${\tilde \rho^{\mbox{\footnotesize{e}},\,\mbox{\footnotesize{o}}}(\bar \omega)}/{\bar \omega}$ at $\bar \omega =0$ within the TR approach. For the $eB=0$ case we subtract the UV contribution. Finally using Eq.~(\ref{eq:limit}) we calculate the electromagnetic conductivity. 

In the reconstruction procedure one has to choose the value of the  regularizer $\gamma$. We found that in the region $\gamma<1$ the width of the resolution function is $\sim 3 T$, but the method becomes unstable what leads to large uncertainties of the calculation. At the same time in the region $\gamma>10$ the method is stable with small uncertainties but the resolution function is rather wide (width $> 4 T$). This, in the region $1<\gamma<10$ the method is stable and the resolution is sufficiently narrow $\sim 3.5 T$. For this reason we vary the regularizer in the region $\gamma \in (1, 10)$ and use $f(\omega)=\omega$. 

For zero magnetic field we subtract the ultraviolet contributions from even and odd spectral densities. 
To this end, we use the model for the spectral densities at large frequencies. Taking into account asymptotic freedom in QCD it is reasonable to assume that real spectral densities at $\omega \gg \Lambda_{QCD}$ do not deviate considerably from their tree level expressions. This assumption will be confirmed below. It allows us to propose the following forms of the spectral densities at large frequencies
\beq
\rho^{\tiny\mbox{e}, \mbox{o}}_{\footnotesize\mbox{\tiny UV}}(\omega)= Z_{\tiny\mbox{e},\mbox{o}}\frac{3}{4\pi^2}\omega^2\tanh \left(\frac{\omega\beta}{4}\right),
\label{eq:ultraviolet_tree}
\eeq
where $Z_{\tiny \mbox{e}},\, Z_{\tiny \mbox{o}}$ are the coefficients for the even and odd branches. At the tree level approximation $Z_{\tiny \mbox{e}} = \frac{1}{2}, Z_{\tiny \mbox{o}} = \frac{3}{2}$, but these coefficients can be renormalized by the interactions. In~\cite{Astrakhantsev:2018oue,astrakhantsev2017temperature} it was shown that the BG method with proper scaling can be used to determine the UV coefficient such as $Z_{\tiny \mbox{e}},\, Z_{\tiny \mbox{o}}$. Following~\cite{Astrakhantsev:2018oue,astrakhantsev2017temperature}, we apply BG approach with the rescaling function\footnote{Notice that in order to account discretization uncertainties we use lattice expressions for the function $f(\omega)$. } $f(\omega) = \frac{3}{4\pi^2}\omega^2\tanh \left(\frac{\omega\beta}{4}\right)$ and use the lattice data for the correlators calculated on the lattice $96\times48^3$.

\begin{figure}[h!]
    \centering
    \includegraphics[scale=0.5]{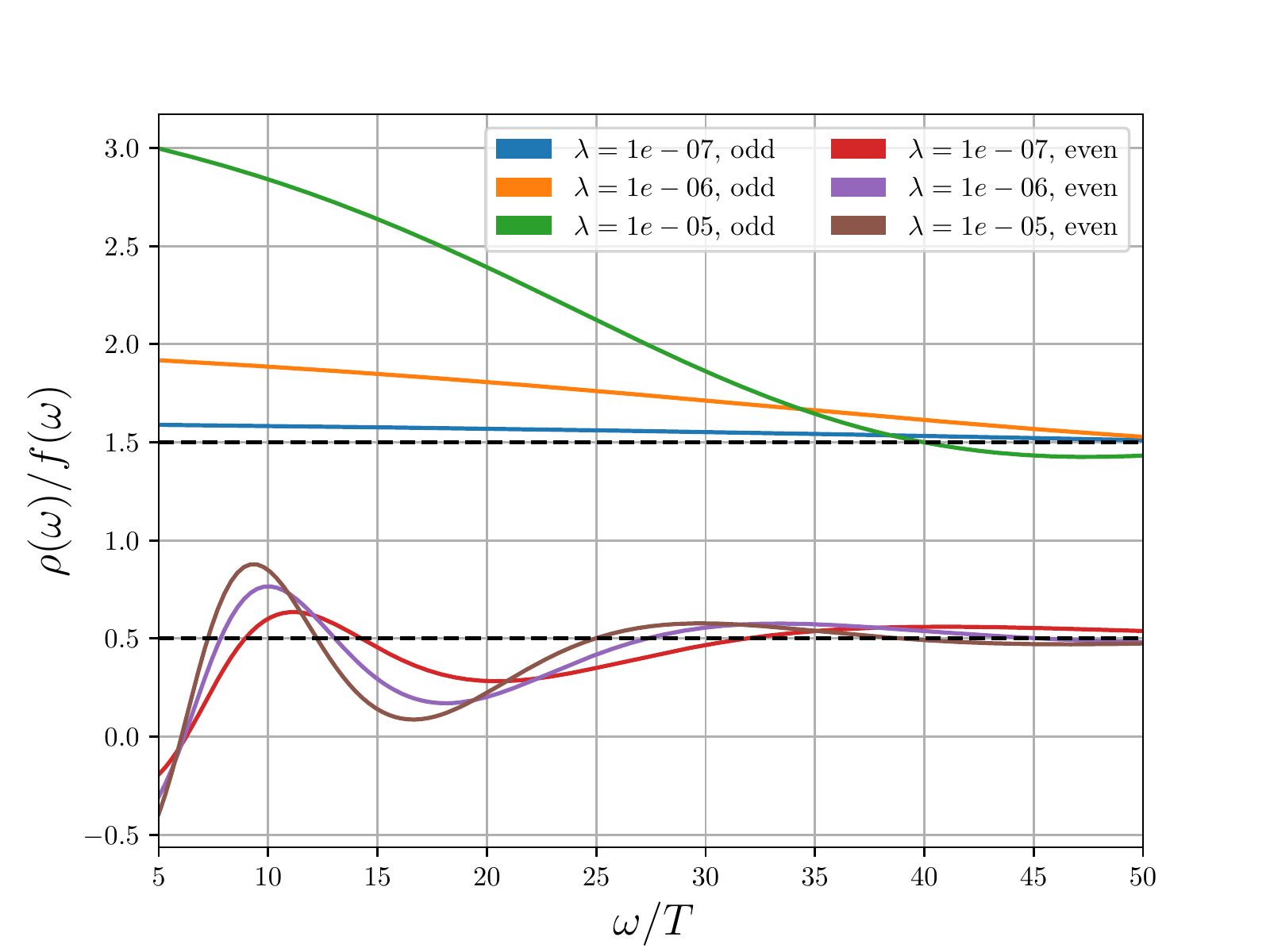}
    \caption{The reconstructed ultraviolet behavior for odd and even branches of the spectral function on the lattice $96\times48^3$ at the tree-level approximation and $f(\omega) = \frac{3}{4\pi^2}\omega^2\tanh \left(\frac{\omega\beta}{4}\right)$.  The reconstruction is carried out for the following values of the  $\lambda = 10^{-5},\,10^{-6},\,10^{-7}$. The dashed lines correspond to tree level results $Z_{\tiny \mbox{e}} = \frac{1}{2}, Z_{\tiny \mbox{o}} = \frac{3}{2}$.}
    \label{fig:UV_free}
\end{figure}

\begin{figure}[h!]
    \centering
    \includegraphics[scale=0.5]{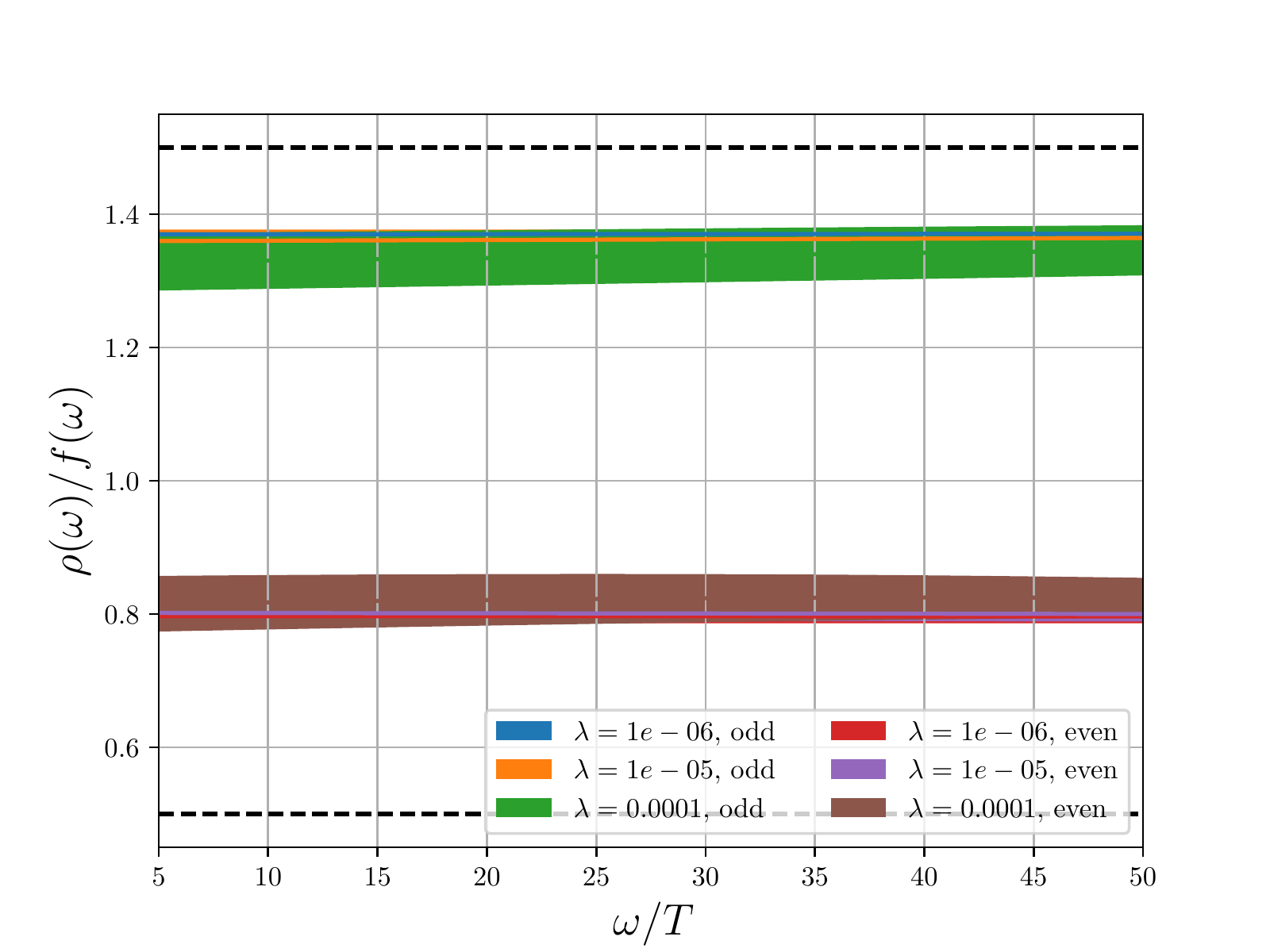}
    \caption{
    The reconstructed ultraviolet behavior for odd and even branches of the spectral function on the lattice $96\times48^3$ in the interacting case and $f(\omega) = \frac{3}{4\pi^2}\omega^2\tanh \left(\frac{\omega\beta}{4}\right)$.  The reconstruction is carried out for the following values of the  $\lambda = 10^{-5},\,10^{-6},\,10^{-7}$. The dashed lines correspond to tree-level results $Z_{\tiny \mbox{e}} = \frac{1}{2}, Z_{\tiny \mbox{o}} = \frac{3}{2}$.}
    \label{fig:UV_int}
\end{figure}

In Fig.~\ref{fig:UV_free} we plot the UV behavior of the rescaled reconstructed spectral function $\tilde \rho(\bar \omega)$ for different values of $\lambda = 10^{-4},\,10^{-5},\,10^{-6}$ at the tree level case. From Fig.~\ref{fig:UV_free} one can see that the reconstructed asymptotic values match the tree-level values $3/2$, $1/2$, what confirms the validity of the method.

In Fig.~\ref{fig:UV_int} we perform the same procedure in the interacting case for $\lambda = 10^{-4},\,10^{-5},\,10^{-6}$. From Fig.~\ref{fig:UV_int} it is seen that in the UV the spectral functions indeed correspond to the models~(\ref{eq:ultraviolet_tree}). 
Notice also that the coefficients $Z_{\tiny \mbox{e}},\, Z_{\tiny \mbox{o}}$ are considerably renormalized as compared to their tree level values, but their mean value is only slightly renormalized, $A = (Z_{\tiny \mbox{e}}+ Z_{\tiny \mbox{o}}) / 2 \sim 1.0$. Taking into account the uncertainties of the calculation we obtain $A=1.05 \pm 0.05$. 
Notice that this result agrees with previous calculations~\cite{Brandt:2015aqk, Ding:2016hua}, where the renormalization of $A$ was shown to be small. 

Finally, to perform the subtraction of the UV contribution, we take the mean value of two branches conductivity $\rho(\omega) = (\rho^e(\omega) + \rho^o(\omega)) / 2$. Then the UV contribution is subtracted in the form 
\begin{equation}
	\Delta \tilde \rho(\bar \omega) = A \int\limits_{\omega_0}^{+\infty} d \omega \delta(\bar \omega, \omega) \frac{3}{4\pi^2}\omega^2\tanh \left(\frac{\omega\beta}{4}\right),
	\label{eq:subtraction}
\end{equation}
where the $\omega_0$ is the frequency which represents the asymptotic freedom region~(\ref{eq:subtraction}) onset. Unfortunately we are not able to determine the value of the $\omega_0$ within the BG method. In the calculation of the conductivity we vary $\omega_0 \in (1.5\,\mbox{GeV}, 3.0\,\mbox{GeV})$ and account this as the systematic uncertainty. Note that this range of $\omega_0$ is in good agreement with the one obtained in~\cite{Brandt:2015aqk} within the fitting procedure. {Using formula (\ref{eq:subtraction})
it is not difficult to find that before the subtraction the UV contribution gives 20-30\% to the reconstructed conductivity.}

\appendix
\subsection{Appendix C: Numerical Setup for the Monte-Carlo Simulations}
\label{appendixC}

We simulate $2+1$ flavours QCD using stout improved rooted staggered fermions and the tree-level Symanzik improved gauge action~\cite{Weisz:1982zw, Curci:1983an}. The partition function is written as 
\begin{equation}\label{eq:partfunc}
Z(B) = \int \!\mathcal{D}U \,e^{-S_{Y\!M}}
\!\!\!\!\prod_{f=u,\,d,\,s} \!\!\!
\det{({D^{f}_{\textnormal{st}}[B]})^{1/4}}\ ,
\end{equation}
where
\begin{equation}\label{eq:tlsyact}
S_{Y\!M}= - \frac{\beta}{3}\sum_{i, \mu \neq \nu} \left(
\frac{5}{6} P^{1\!\times \! 1}_{i;\,\mu\nu} - \frac{1}{12}
P^{1\!\times \! 2}_{i;\,\mu\nu} \right)\ ,
\end{equation}
and the symbols $P^{1\!\times \! 1}_{i;\,\mu\nu}$ and $P^{1\!\times \!
2}_{i;\,\mu\nu}$ denote the real part of the trace of $1\!\times \!
1$ and $1\!\times \!2$ loops. The staggered matrix is
\begin{equation}\label{eq:rmmatrix}
\begin{aligned}
(D^f_{\textnormal{st}})_{i,\,j} =\ & am_f
  \delta_{i,\,j}+\!\!\sum_{\nu=1}^{4}\frac{\eta_{i;\,\nu}}{2}
  \left(u^f_{i;\,\nu}U^{(2)}_{i;\,\nu}\delta_{i,j-\hat{\nu}} \right. \\ 
  &-\left. u^{f*}_{i-\hat\nu;\,\nu}U^{(2)\dagger}_{i-\hat\nu;\,\nu}\delta_{i,j+\hat\nu}
  \right)\ , 
\end{aligned}
\end{equation}
where $\eta_{i;\,\nu}$s are the staggered phases, $U^{(2)}_{i;\,\mu}$ stands for the two times stout-smeared link~\cite{Morningstar:2003gk} (with isotropic smearing parameter $\rho=0.15$) and $u^f_{i;\,\mu}$ is the abelian field phase. 

The abelian transporters corresponding to a uniform magnetic field $B_z$ directed along $\hat{z}$ can be chosen as
\begin{eqnarray}\label{eq:bfield}
u^f_{i;\,y}=e^{i a^2 q_f B_{z} i_x} \ , \quad
{u^f_{i;\,x}|}_{i_x=N_x}=e^{-ia^2 q_f N_x B_z i_y}\, ,
\end{eqnarray}
where $q_f$ is the quark charge and all the other abelian links are set to 1 ($N_k$ is the lattice extent in the $\hat{k}$ direction, $1\le i_k\le N_k$).  $B_z$ cannot be arbitrary: for Eq.~\eqref{eq:bfield} to describe a uniform magnetic field on a lattice torus, the value $B_z$ must be quantized as follows~\cite{tHooft:1979rtg, Damgaard:1988hh, AlHashimi:2008hr}
\begin{equation}\label{bquant}
\frac{e}{3}B_z={2 \pi b}/{(a^2 N_x N_y)}\ ,
\end{equation}
where $b$ is an integer. 

Bare parameters have been chosen so as to stay on a line of constant physics with physical quark masses. In particular, we adopted the values reported in~\cite{Aoki:2009sc,Borsanyi:2010cj, Borsanyi:2013bia}, either directly or by interpolation. $O(100)$ decorrelated gauge configurations have been used for each simulation point. Gauge configurations have been sampled using the Rational Hybrid Monte-Carlo algorithm~\cite{Clark:2004cp, Clark:2006fx,Clark:2006wp}.

\bibliography{main}

\end{document}